\documentclass[manuscript]{aastex}
\usepackage{multirow}

\slugcomment{}

\shorttitle{Molecular clouds in the Trifid nebula M20}
\shortauthors{K. Torii et al.}

\begin{document}


\title{Molecular clouds in the Trifid nebula M20; \\
 Possible evidence for a cloud-cloud collision in triggering the formation of the first generation stars}


\author{K. Torii\altaffilmark{1}, R. Enokiya\altaffilmark{1}, H. Sano\altaffilmark{1}, S. Yoshiike\altaffilmark{1}, N. Hanakoka\altaffilmark{1}, A. Ohama\altaffilmark{1}, N. Furukawa\altaffilmark{1},  J. R. Dawson\altaffilmark{1,2}, N. Moribe\altaffilmark{1}, K. Oishi\altaffilmark{1}, Y. Nakashima\altaffilmark{1}, T. Okuda\altaffilmark{1}, H. Yamamoto\altaffilmark{1} A. Kawamura\altaffilmark{1,3}, N. Mizuno\altaffilmark{1,3}, H.Maezawa\altaffilmark{4,5}, T. Onishi\altaffilmark{1,4}, A.Mizuno\altaffilmark{5} and Y. Fukui\altaffilmark{1}}
\affil{$^1$Department of Physics and Astrophysics, Nagoya University, Chikusa-ku, Nagoya, Aichi, 464-8601, Japan}
\affil{$^2$School of Maths and Physics, University of Tasmania, Private Bag 37, Hobart Tasmania, 7001, Australia}
\affil{$^3$National Astronomical Observatory of Japan, Mitaka, Tokyo, 181-8588, Japan}
\affil{$^4$Department of Astrophysics, Graduate School of Science, Osaka Prefecture University, 1-1 Gakuen-cho, Nakaku, Sakai, Osaka 599-8531, Japan}
\affil{$^5$Solar-Terrestrial Environment Laboratory, Nagoya University, Chikusa-ku, Nagoya 464-8601, Japan}

\email{torii@a.phys.nagoya-u.ac.jp}




\begin{abstract}

A large-scale study of the molecular clouds toward the Trifid nebula, M20, has been made in the $J$=2--1 and $J$=1--0 transitions of $^{12}$CO and $^{13}$CO. M20 is ionized predominantly by an O7.5 star HD164492.  The study has revealed that there are two molecular components at separate velocities peaked toward the center of M20 and that their temperatures --- 30--50 K as derived by an LVG analysis --- are significantly higher than the 10 K of their surroundings. We identify that the two clouds as the parent clouds of the first generation stars in M20. The mass of each cloud is estimated to be $\sim10^3$ $M_\odot$ and their separation velocity is $\sim8$ km s$^{-1}$ over $\sim$1--2 pc. We find the total mass of stars and molecular gas in M20 is less than $\sim3.2\times10^3$ $M_\odot$, which is too small by an order of magnitude to gravitationally bind the system. We argue that the formation of the first generation stars, including the main ionizing O7.5 s
 tar, was triggered by the collision between the two clouds in a short time scale of $\sim$1 Myrs, a second example alongside Westerlund 2, where a super star cluster may have been formed due to cloud-cloud collision triggering.

\end{abstract}





\section{Introduction}

High-mass stars play an important role in dynamically agitating and ionizing the interstellar medium in galaxies and strongly influence galactic evolution. It is therefore crucial to understand their formation mechanisms. Despite numerous theoretical and observational studies attempting to address this issue in the last few decades, the detailed physical processes involved are still elusive \citep[e.g.,][]{zin2007}. Since high-mass stars generally form in clusters, addressing the formation of the host cluster is a more fundamental process in the study of high-mass star formation, and one which would benefit from more attention.

External effects, such as multiple supernova explosions that form supershells \citep[e.g.,][]{daw2011} and HII regions \citep[e.g.,][]{yam2001}, may play an important role in the formation of clusters, because compressive triggering increases the external pressure, promoting massive cluster formation. Similarly, in the Antennae galaxies, a galaxy-galaxy collision played a role in triggering cluster formation on kpc scales, and several young globular clusters may have been formed \citep{wil2000}. The collision must have created strong shocks with a relative speed of $\sim$100 km s$^{-1}$ or more,  which presumably lead to the rapid formation of the natal massive and compact pre-cluster clouds. The distance of these galaxies $\sim$20 Mpc, however, does not allow one to resolve the distribution of molecular gas toward individual clusters, and the detailed mechanism of cluster formation here remains as an open issue. 

 More recently, \citet{fur2009} discovered two giant molecular clouds (GMCs) toward the young cluster Westerlund 2 that are likely to have recently collided with each other. Westerlund 2 is a remarkable Galactic super-star cluster which includes more than 10 high-mass stars with total stellar mass of $\sim$4500 $M_\odot$ in a small volume only $\sim$1 pc in radius, and which has an age of a few Myrs and is located at a distance of 3--8 kpc in the Galaxy \citep{asc2007,fur2009, rau2007}. \citet{fur2009} presented morphological evidence for the physical association of the clouds with the cluster and its ionizing HII region RCW49 and argued that the cluster may have been formed via a collision between the two GMCs as suggested by their relative positions, masses and velocities. Subsequently, \citet{oha2010} derived significant temperature enhancements in the GMCs toward RCW49, offering further robust verification of their association with Westerlund 2. \citet{hig2010} have also presented a study of 14 young star clusters associated with dense molecular clumps of several hundreds of solar masses, and suggested clump-clump collisions as a possible formation mechanism for some of these objects. However, studies of the cloud-cloud collision scenario are yet limited to a small number of clusters and it is important to find more collision-triggered cluster formation in order to better understand the formation of clusters and high-mass stars.

 The Trifid Nebula (M20, NGC6514) is a well-known open cluster with three outstanding dust lanes \citep{wal1973,cha1975}, with an estimated distance from 1.7 kpc \citep{lyn1985} to 2.7 kpc \citep{cam2011}.  An optical study detected 320 stars within 10 pc of the center \citep{ogu1975}, and a near-infrared study identified 85 T-Tauri stars in the nebula \citep{rho2001}.  The total stellar mass of M20 is estimated to be less than 500 $M_\odot$, several times less massive than Westerlund 2. M20 also harbors one of the youngest known clusters with an estimated age of $\sim$0.3 Myr \citep{cer1998}. Ionization of the HII region, which extends by $\sim$3--5 pc in diameter, is dominated by an O7.5 star HD164492, and some of the low-mass stars in the cluster appear to have been formed together with this O star as suggested by their spatial proximity of a few pc \citep[e.g.,][]{rho2008}. The mass-loss rate of the O-star and expansion velocity of the stellar wind are estimated to be $\dot{M}\sim2\times10^{-6}$ $M_\odot$ yr$^{-1}$ \citep{how1989} and $V_\infty\sim1582$ km s$^{-1}$ \citep{pri1990}, respectively, and the total dynamical luminosity of the wind is calculated to be $1/2\dot{M}V_\infty^2 \sim 1.6\times10^{36}$ erg s$^{-1}$.   In addition, M20 is an on-going star forming region.  Observations at multiple wavelengths revealed evidence for second generation star formation triggered by the first generation stars and for related activities including optical jets \citep[e.g., HH399, ][]{cer1998,hes2004} mid- and far-infrared protostellar objects \citep{rho2006}, infrared and X-ray young stellar objects \citep{rho2001,rho2004, lef2001}, H$\alpha$ emission stars \citep{her1957,yus2005}, etc.  M20 therefore offers a good opportunity for a comprehensive study of cluster formation; from the first generation O-star and many low mass stars to the subsequent second generation young stellar objects. 
 
  Detecting and characterizing any residual natal molecular gas is crucial for understanding the formation of clusters.   Detailed studies of molecular line emission in M20 were presented by \citet{cer1998} and extended by \citet{lef2000} and \citet{lef2008}.  They covered a $20' \times 20'$ area centered on the O7.5 star HD164492 with a high angular resolution of few tens of arcseconds and detected molecular structures apparently associated with M20 in the velocity range 0--30 km s$^{-1}$. Parts of the molecular features apparently trace the three dark lanes.  They also identified cold dust cores around M20 from observations of millimeter continuum emission and discussed on-going star formation, with candidates for young stellar objects (YSO) detected by the Spitzer telescope \citep{cer1998,lef2008,rho2006}.  However, it is not yet understood how these molecular clouds are organized on a scale covering the whole cluster, and the possiblility of triggered formation of the cluster has not been explored.

We present here new CO $J$=2--1 and $J$=1--0 observations of M20 with the NANTEN2/NANTEN 4m sub-mm/mm telescope in Chile. Our principle aim is to reveal degree-scale distribution of molecular gas in order to investigate the formation mechanisms of the first generation stars in M20, especially the central O star and the associated cluster. This paper is organized as follows; Section 2 summarizes the observations and Section 3 the results. Discussion is given in Section 4 and a summary in Section 5.

\section{Observations}

Observations of the $J$=2--1 transition of CO were made with the NANTEN2 4 m sub-millimeter telescope of Nagoya University at Atacama (4865 m above the sea level) in Chile in January-February and October-November 2008 for $^{12}$CO and in April-November 2008 for $^{13}$CO. The half-power beam width (HPBW) of the telescope was 90$''$ at 230 GHz. The 4 K cooled SIS mixer receiver provided a typical system temperature of $\sim$200 K in the single-side band at 220--230 GHz, including the atmosphere toward the zenith. The spectrometer was an acousto-optical spectrometer with 2048 channels, providing a velocity coverage of 392 km s$^{-1}$ with a velocity resolution of 0.38 km s$^{-1}$ at 230 GHz. We observed a large area of M20 and the supernovae remnant (SNR) W28 in $^{12}$CO($J$=2--1), while $^{13}$CO($J$=2--1) observations are limited in a smaller area as shown in Figure \ref{largemap}. The OTF (= on-the-fly) mapping mode was used in the observations, and the output grid of the 
 observations is 30$''$. We smoothed the velocity resolution and spatial resolution to 0.95 km s$^{-1}$ and 100$''$, respectively, to achieve a better noise level. Finally, we obtained the rms noise fluctuations of $\sim$0.3 K and $\sim$0.1 K  per channel in $^{12}$CO and $^{13}$CO, respectively.

Observations of the $^{12}$CO($J$=1--0) and $^{13}$CO($J$=1--0) transitions were made with the NANTEN 4m millimeter telescope of Nagoya University at the Las Campanas Observatory in Chile and were carried out with a 4 K cryogenically cooled Nb superconductor-insulator-superconductor (SIS) mixer receiver \citep{oga1990} during the period from March 1999 to September 2001 for $^{12}$CO($J$=1--0) and in October 2003 for $^{13}$CO($J$=1--0). The spectrometer was an acousto-optical spectrometer (AOS) with 2048 channels, which gave a frequency band-width and resolution of 250 MHz and 250 kHz, corresponding to a velocity coverage of 650 km s$^{-1}$ and a velocity resolution of 0.65 km s$^{-1}$ at 115 GHz. The HPBW of the telescope was 2$'$.6. Observations were made in the position-switching mode with a 4$'$ grid spacing. The typical system temperature was $\sim$280 K in the single-side band, and typical rms noise level was 0.36 K at a velocity resolution of 0.65 km s$^{-1}$. 

\section{Results}
\subsection{Observational results }

\begin{figure}
\epsscale{.80}
\plotone{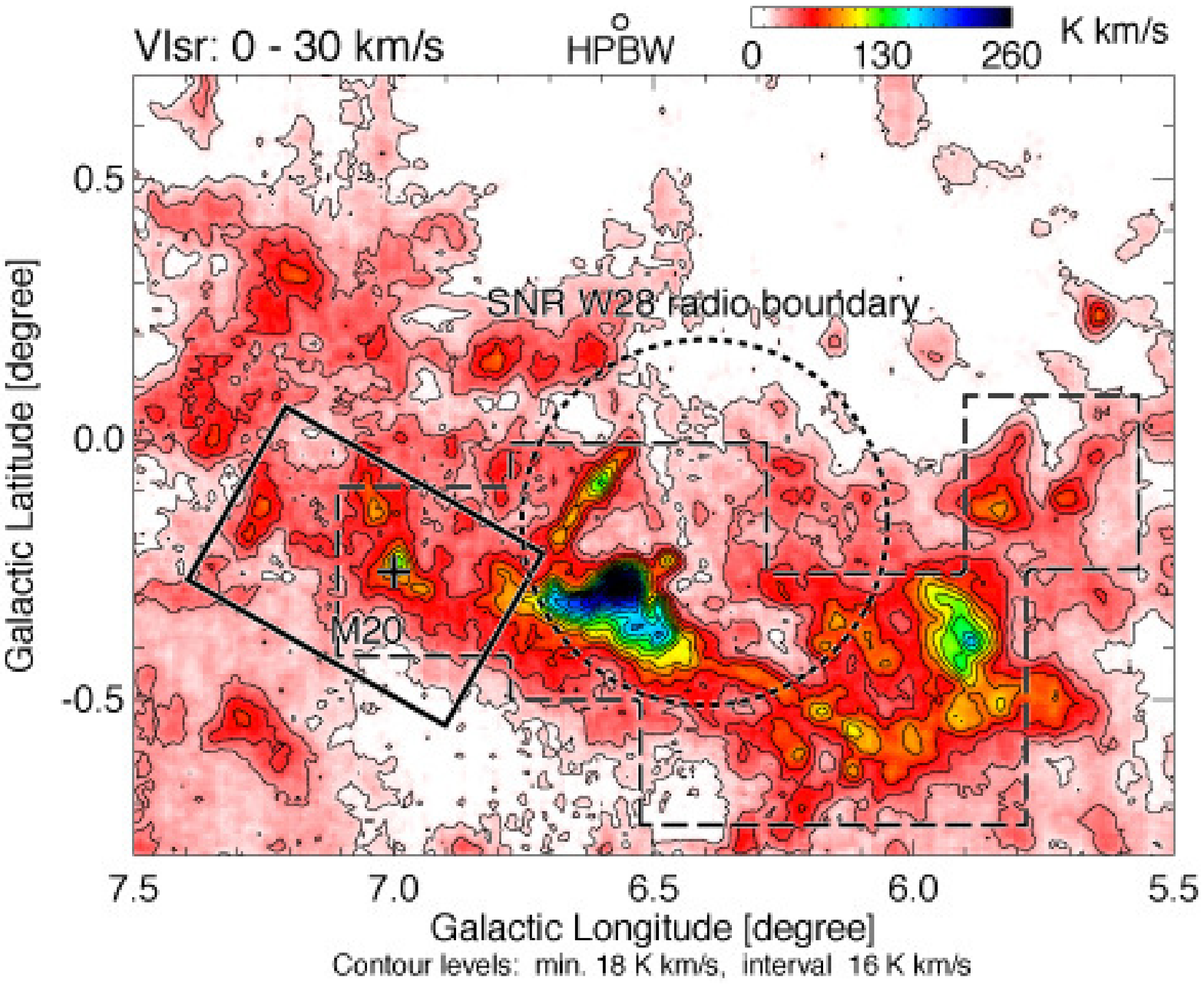}
\caption{Integrated intensity distribution of $^{12}$CO($J$=2--1) emission towards ($l$, $b$) $\sim$ (5.5$^\circ$--7.5$^\circ$, $-$0.8$^\circ$--0.7$^\circ$). Contours are drawn every 16 K km s$^{-1}$ from 18 K km s$^{-1}$. The cross indicates M20, and solid lines indicates the region shown in Figure \ref{iimap}. The dotted circle indicates the boundary of radio emission from W28, and dashed lines indicate the region observed in $^{13}$CO($J$=2--1).\label{largemap}}
\end{figure}

Figure \ref{largemap} shows the $^{12}$CO($J$=1--0) integrated intensity distributions between $V_{\rm{lsr}}$ of 0 and 30 km s$^{-1}$ in the region $(l, b) \sim (5^\circ.5$--$7^\circ.5, -0^\circ.8$--$0^\circ.7)$. M20 is denoted by a cross and is located at $\sim0^\circ.6$ from the supernova remnant (SNR) W28. The distribution of molecular clouds is very complicated, and includes an elongated feature running from south-west to north-east throughout the region. Dense molecular clouds associated with W28 are found around $(l,b) \sim (6^\circ.4$--$6^\circ.8, -0^\circ.4$--$0^\circ.0)$ in this velocity range \citep[e.g.,][]{ari1999}.

\begin{figure}
\epsscale{.75}
\plotone{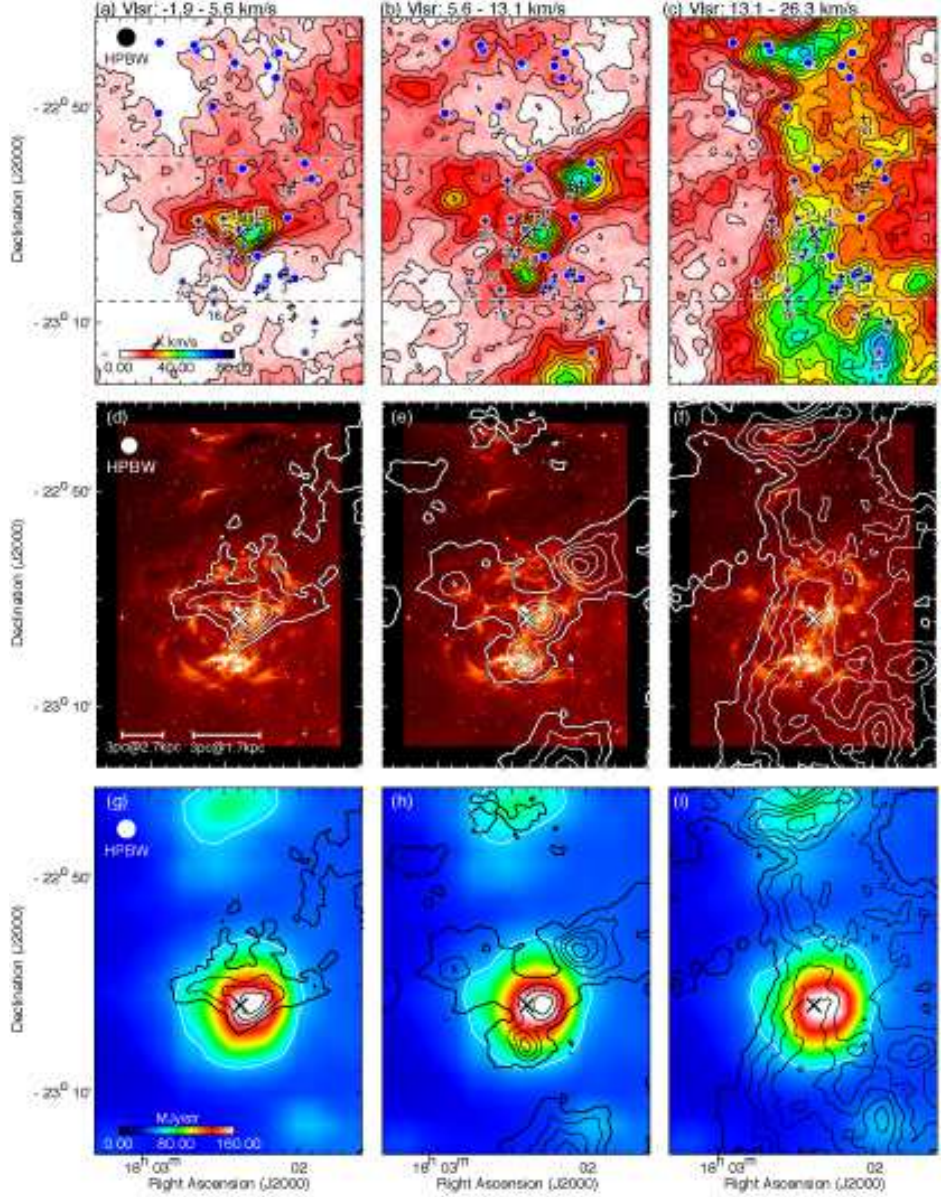}
\caption{(a--c) Integrated intensity distribution of $^{12}$CO($J$=2--1) emission. Contours are drawn every 4 K km s$^{-1}$ from 6 K km s$^{-1}$. Filled circles depict Class 0/I objects identified by Spitzer observations \citep{rho2008}, and small crosses depict cold dust cores (TC00--TC17) identified by radio observations by \citet{lef2008}. The large cross depicts the central O star. (d--f) Contour maps of the $^{12}$CO($J$=2--1) emission superposed on the Spitzer 8 $\mu$m image of \citet{rho2006}. Contours are drawn every 8 K km s$^{-1}$ from 14 K km s$^{-1}$. (g--i) Contour maps of the $^{12}$CO($J$=2--1) emission superposed on IRAS 25 $\mu$m data. White contours mark the two-third peak intensity of the IRAS emission equal to 60 MJy str$^{-1}$. Solid contours show $^{12}$CO($J$=2--1) emission and are drawn every 8 K km s$^{-1}$ from 14 K km s$^{-1}$.\label{iimap}}
\end{figure}

The region shown in Figure \ref{iimap} corresponds to the box drawn in solid lines in Figure \ref{largemap}. Figures \ref{iimap}a--\ref{iimap}c show the integrated intensity distributions of $^{12}$CO($J$=2--1) over three different velocity ranges, Figures \ref{iimap}d--\ref{iimap}f show comparisons between the $^{12}$CO($J$=2--1) and the Spitzer 8¦Ìm data, and Figures \ref{iimap}g--\ref{iimap}i show comparisons between the $^{12}$CO($J$=2--1)  and the IRAS 25 $\mu$m data. The large cross depicts the central star. Class 0/I objects and cold dust cores \citep[cloud cores TC00--TC17][]{lef2008} are shown by circles and small crosses, respectively. These objects indicate recent star formation and pre-star formation of the second generation of stars in M20. 

  In Figures \ref{iimap}a, \ref{iimap}d and \ref{iimap}g, we find a cloud with its peak position near the central star and well aligned with the center of the Trifid dust lanes (hereafter 2 km s$^{-1}$ cloud). The distribution of the molecular gas shows a triangle shape and each of its corners appears to trace the three dust lanes.

\begin{figure}
\epsscale{.75}
\plotone{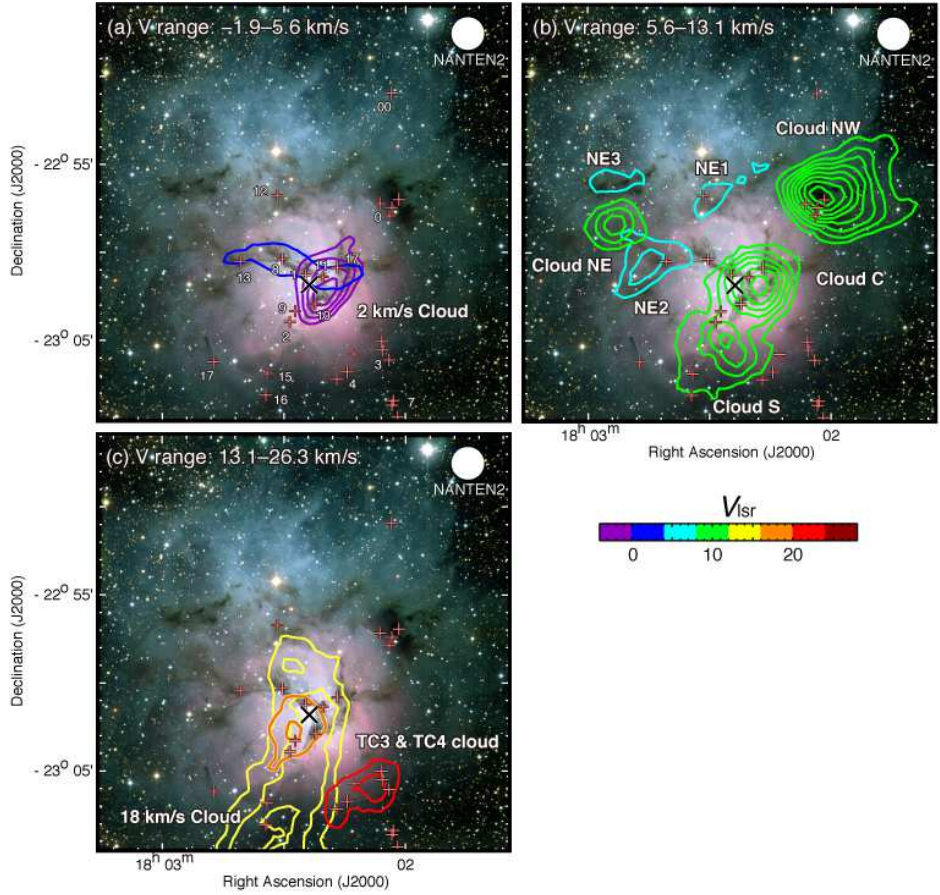}
\caption{CO($J$=2--1) distribution of the molecular clouds associated with M20, superposed on an optical image (Credit: NOAO). Contours for all clouds are drawn every 4 K km s$^{-1}$ from 14 K km s$^{-1}$. Small crosses depict cold dust cores, with ID numbers labeled only in Figure \ref{co+noao}a. The large crosses depict the central star. \label{co+noao}}
\end{figure}

The molecular gas around 8 km s$^{-1}$, shown in Figures \ref{iimap}b, \ref{iimap}e and \ref{iimap}h, consists of a central cloud and three surrounding clouds to the northeast, northwest and south of (hereafter clouds C, NW, NE and S, respectively). In addition, we list three small clouds named clouds NE1, NE2 and NE3, respectively, as indicated in Figure \ref{co+noao}b. The peak position of cloud C corresponds to that of the 2 km s$^{-1}$ cloud. Cloud NW is associated with TC0 without any Class 0/I objects. Cloud S are located at south of cloud C with TC2, and Cloud NE appears to be located at the top of a molecular filament extending to the east from cloud C. 

The molecular gas around $V_{\rm{lsr}} \sim 18$ km s$^{-1}$ shown in Figures \ref{iimap}c, \ref{iimap}f and \ref{iimap}i extends from the north to the south throughout the region, and we found that there is a elongated molecular feature overlapping with the central star on its north at  $(Ra, Dec) \sim (18^h02^m10^s$--$18^h02^m40^s$, $-23^\circ16'$--$-22^\circ46')$ (hereafter, 18 km s$^{-1}$). Some of the other molecular components in this velocity range are likely associated with a infrared dark filament in M20. \citet{cam2011} studied the extinction towards M20 and found that cold dust cores TC00, TC0, TC3, TC4 and TC5 are located toward the dark filaments (Figure \ref{co+noao}). These cold cores show molecular counterparts in this velocity range. A cloud containing TC3, TC4 and several Class 0/I objects are distributed at $(Ra,Dec) \sim (18^h2^m6^s, -23^\circ6'0'')$, just west of the 18 km s$^{-1}$ cloud (hereafter TC3 \& TC4 cloud). We also found molecular clouds associated with the Trifid Junior, which is another infrared nebula at $Dec \sim -22^\circ46'$--$-22^\circ50'$ detected by Spitzer \citep{rho2008}. The total integrated intensity of $^{12}$CO($J$=1--0) in this velocity range accounts for about 60 \% of the sum of those in three velocity ranges, meaning that a large fraction of molecular mass in this region is concentrated to around $V_{\rm{lsr}}$ of $\sim$18 km s$^{-1}$.

The detailed distribution of the molecular gas shows good positional coincidence with optical dark features (Figure \ref{co+noao}). The velocity ranges of the three panels in Figure \ref{co+noao} correspond to those of Figures \ref{iimap}a, b and c. The current low angular resolution is not sufficient to resolve individual correspondences. However, the general tend may still be discerned in Figure \ref{co+noao}. We find that many molecular features correspond well to the dark lanes and dark clouds in the optical image, indicating that they are located on the front side of, or partially embedded in, the nebula.

Figure \ref{co+noao}a shows that the 2 km s$^{-1}$ cloud is divided into two components and each of them trace the four major dust lanes elongated from the center to the east, the west, the north and the southeast. Cold dust cores TC1, TC11, TC8, TC10 and TC13 are associated with this cloud (Lefloch et al. 2008). Figure \ref{co+noao}b shows that cloud C coincides with the center of the nebula whereas it does not match well the dark lanes, suggesting that cloud C is located within or on the far side of the nebula. Cloud S seems to correspond to the pillar-like feature in the south of the nebula associated with TC2 \citep{lef2002}, indicating that it is located within or on the far side of the nebula. Cloud NW, which has the highest $^{12}$CO intensity, corresponds well to a dark feature in the northwest, and interactions between cloud NW and the HII region have been suggested by \citet{cer1998} and \citet{lef2008}. Two small features in the north, clouds NE1 and NE3, correspond to small dark lanes and cloud NE2 corresponds to the eastern extension of the 2 km s$^{-1}$ cloud along the major dark lane. Figure \ref{co+noao}c shows that the 18 km s$^{-1}$ cloud has a peak component to the southwest of the central star and \cite{lef2008} indicates that TC9 is associated with the cloud. The 18 km s$^{-1}$ cloud has no counterpart dark features in the optical image and is likely located on the far side of M20. The total $A_{V}$ in the vicinity of the central star is 10--15 mag \citep{cam2011}, while a direct measurement toward the central star itself gives a quite small $A_{V}$ of 1.3 mag \citep{lyn1985}, indicating that a large fraction of the gas toward the central star is located on the far side of the nebula. We suggest that although the contours of the 2 km s$^{-1}$ cloud and cloud C appear to overlap with the O star, this is likely the result of poor resolution in CO, which spreads the emission over a wider spatial area that it in fact occupies. The high resolution dust image in Figure \ref{iimap}d--f indicates that the dust filaments are located toward the west of the O star with some offset, suggesting that the amount of actual obscuring matter is minimal toward the O star.  This minimal obscuration is consistent with a far-side location for the 18 km s$^{-1}$ cloud. The TC3 \& TC4 cloud is also clearly seen in Figure 3c. \cite{lef2000} and \cite{rho2008} studied this cloud in detail and indicated that it is associated with the HII region at its boundary. 

\begin{figure}
\epsscale{.40}
\plotone{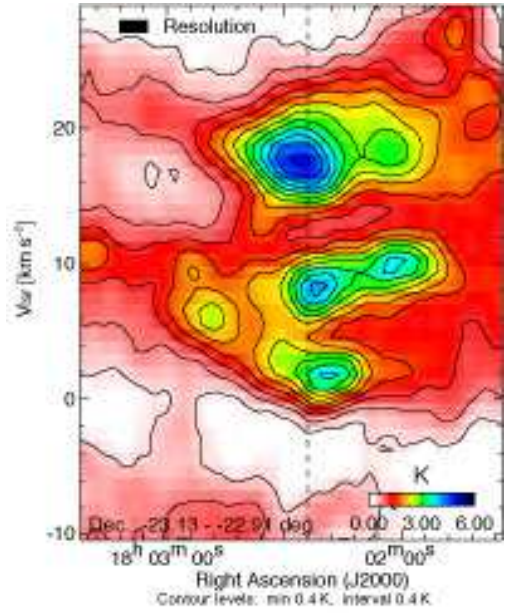}
\caption{Position--velocity diagram of $^{12}$CO($J$=2--1) emission. The integration range is shown in Figure 1(a--c) by dashed lines. The dashed line in the figure shows the position of the central O star of M20.\label{lv}}
\end{figure}

  Figure \ref{lv} shows a position-velocity diagram of the molecular clouds. The integration range in Declination by the dotted lines is shown in Figures \ref{iimap}a--\ref{iimap}c.  Three velocity components are clearly seen. We find that the 2 km s$^{-1}$ cloud has a positive velocity gradient to the east, while the 8 km s$^{-1}$ clouds have a negative velocity gradient to the east, and that these two velocity components seem to be connected with each other at $V_{\rm lsr}\sim$6 km s$^{-1}$ around the position of the O star.  On the other hand, the 18 km s$^{-1}$ clouds are extensively distributed and are detached from the 2 km s$^{-1}$ and 8 km s$^{-1}$ clouds in velocity space. 

\begin{figure}
\epsscale{.80}
\plotone{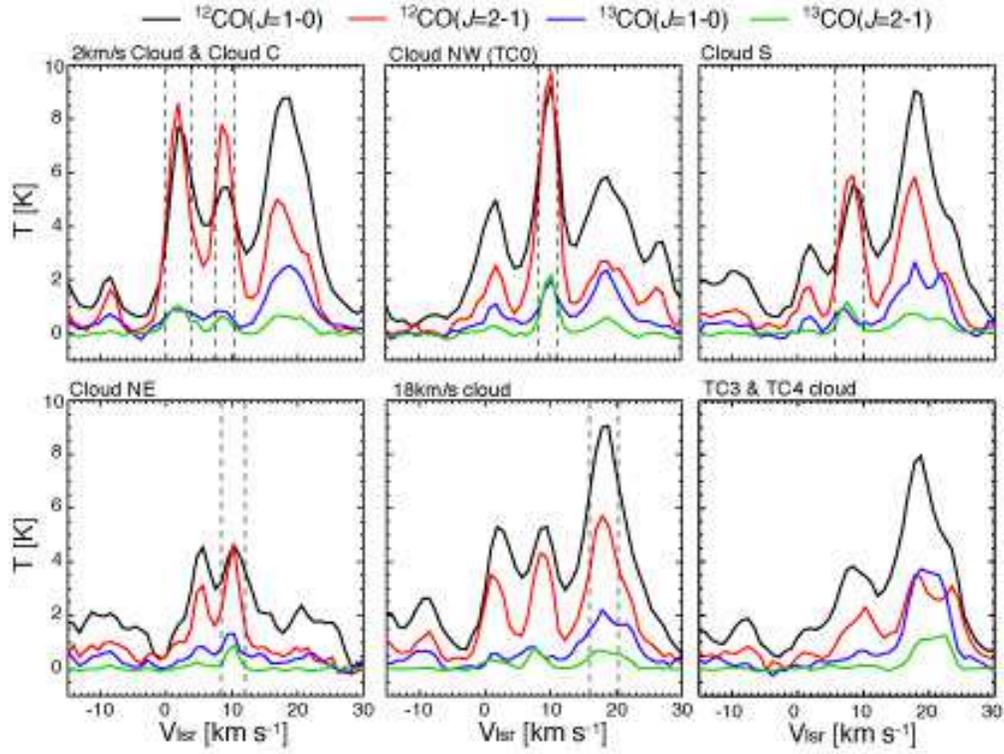}
\caption{CO spectra at the peak positions of the clouds listed in table \ref{cloudlist}. $^{12}$CO($J$=1--0), $^{12}$CO($J$=2--1), $^{13}$CO($J$=1--0) and $^{13}$CO($J$=2--1) are plotted in black, red, blue and green, respectively. All spectra were smoothed to be a beam size of $2.\arcmin6$. Dotted lines indicated the velocity ranges used for the LVG calculations shown in figures \ref{lvg1} and \ref{lvg2}. \label{spec}}
\end{figure}

We define seven clouds, the 2 km s$^{-1}$ cloud, clouds C, NW, S, NE, the 18 km s$^{-1}$ cloud and TC3 \& TC4 cloud, in $^{12}$CO($J$=2--1) by plotting contours at half of the peak integrated intensity shown in Figure \ref{iimap}a--c, and if other peaks are found inside the contour, we distinguish these by plotting a line at the valley of the intensity distributions. (We here use a threshold of two-thirds of the peak intensity for the 18 km s$^{-1}$ cloud and TC3 \& TC4 cloud because the contour at half peak intensity contains too many other peaks.) Figure \ref{spec} shows the observed spectra at the peak positions of the seven clouds. We here assume two heliocentric-distances, 1.7 kpc and 2.7 kpc for estimating sizes and masses of the clouds.  Details of the identified clouds are shown in Table \ref{cloudlist}. Here we can not measure velocity information for the TC3 \& TC4 cloud because of self absorption in the spectrum at a velocity of $\sim$21 km s$^{-1}$. Velocity widths of the spectra at the peaks are $\sim$3--6 km s$^{-1}$, with the 18 km s$^{-1}$ cloud showing a somewhat wider spectrum than the others. The radii of the clouds are estimated to be $\sim$1--4 pc. The masses are estimated to be a $\sim$1--2$\times 10^3 M_\odot$ for the 2 km s$^{-1}$ and 8 km s$^{-1}$ clouds and $\sim$3--4$\times10^3 M_\odot$ for the 18 km s$^{-1}$ cloud. We here assume an X-factor of $2.0\times10^{20}$ cm$^{-2}$ (K km s$^{-1}$)$^{-1}$ \citep{str1988} to convert $^{12}$CO($J$=1--0) integrated intensity into H$_2$ column density in estimating these masses. 
As a check against previous work, we may estimate the entire H$_2$ mass of this region within the full velocity range ($-1.9$--26.3 km s$^{-1}$). This results in a derived mass of 8.5$\times10^4$ M$_\odot$ and 2.2$\times10^5$ M$_\odot$ for distances of 1.7 kpc and 2.7 kpc, respectively.  \cite{cam2011} estimated the mass of a somewhat larger area in this region to be 5.8$\times10^5$ M$_\odot$ by assuming a distance of 2.7 kpc, which is comparable to our estimate.

\begin{figure}
\epsscale{.80}
\plotone{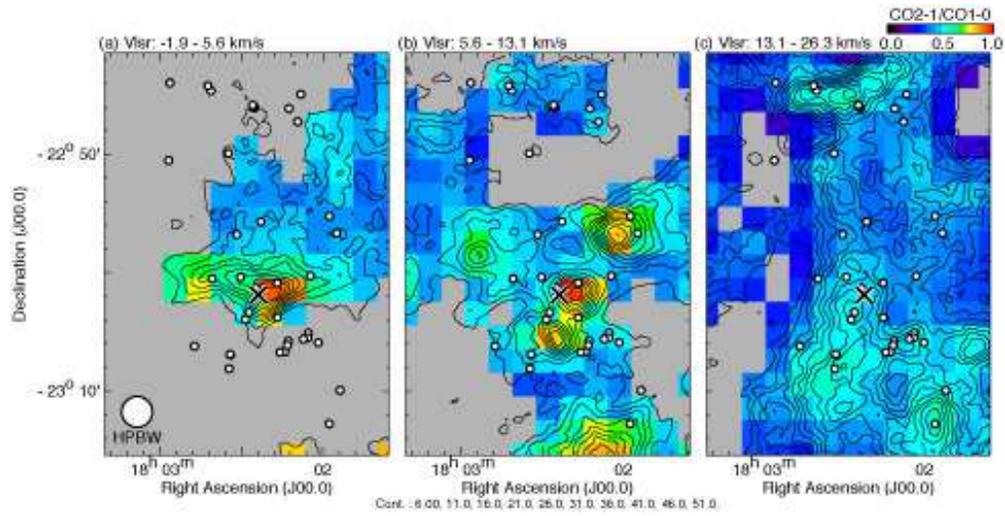}
\caption{Spatial distribution of $^{12}$CO($J$=2--1) emission (contours) and ratio between  $^{12}$CO($J$=2--1) and  $^{12}$CO($J$=1--0). The data used for the calculation of ratios was spatially smoothed with a gaussian function to an effective beam size of 2$'$.6. Contours are plotted at every 5 K km s$^{-1}$ from 11 K km s$^{-1}$. Circles depict Class 0/I objects \citep{rho2008} and the cross depicts the central O star of M20.\label{ratio}}
\end{figure}

Figure \ref{ratio} shows the spatial distribution of the ratio between the $^{12}$CO($J$=1--0) and $^{12}$CO($J$=2--1) transitions, which represents the degree of the rotational excitation of the molecular gas and reflects temperature and density. We find that the 2 km s$^{-1}$ cloud and 8 km s$^{-1}$ clouds C, NW and S show significantly higher ratios of $>$ 0.8, while the cloud NE shows somewhat smaller value of 0.6 at its maximum. Clouds at $V_{\rm{lst}} \sim 18$ km s$^{-1}$ including the 18 km s$^{-1}$ cloud and TC3 \& TC4 cloud show typically smaller ratios of $\sim$0.5 and up to $\sim$0.8. These results are also seen in Figure \ref{spec}. The molecular gas at the southern edge of Figure \ref{iimap}b is associated with the nearby SNR W28 and shows exceptionally higher ratios of $\sim$1.0.

\subsection{Physical parameters of the molecular clouds}

We estimate the temperature and density of the M20 molecular clouds with an LVG analysis \citep[e.g.,][]{gol1974}. We took $^{12}$CO($J$=2--1), $^{13}$CO($J$=1--0) and $^{13}$CO($J$=2--1) line ratios at 6 positions, as shown in Figure \ref{spec} and Table \ref{cloudlist}. We did not use $^{12}$CO($J$=1--0) because this transition is optically thick and traces mainly the lower density envelopes of the molecular clouds, and we did not estimate parameters for the TC3 \& TC4 cloud because it shows self absorption in its $^{12}$CO spectra, which prevent us from measuring accurate line intensities. The LVG approximation assumes a cloud with uniform velocity gradient, which may not always be correct in HII regions. However, \cite{leu1976} and subsequent studies \citep[e.g.,][]{whi1977} investigated radiative transfer calculations with a microturbulent cloud associated with HII regions and concluded that there are not any remarkable differences between the LVG and microturbulence models. We therefore adopt the LVG approximation in the following.

\begin{figure}
\epsscale{.80}
\plotone{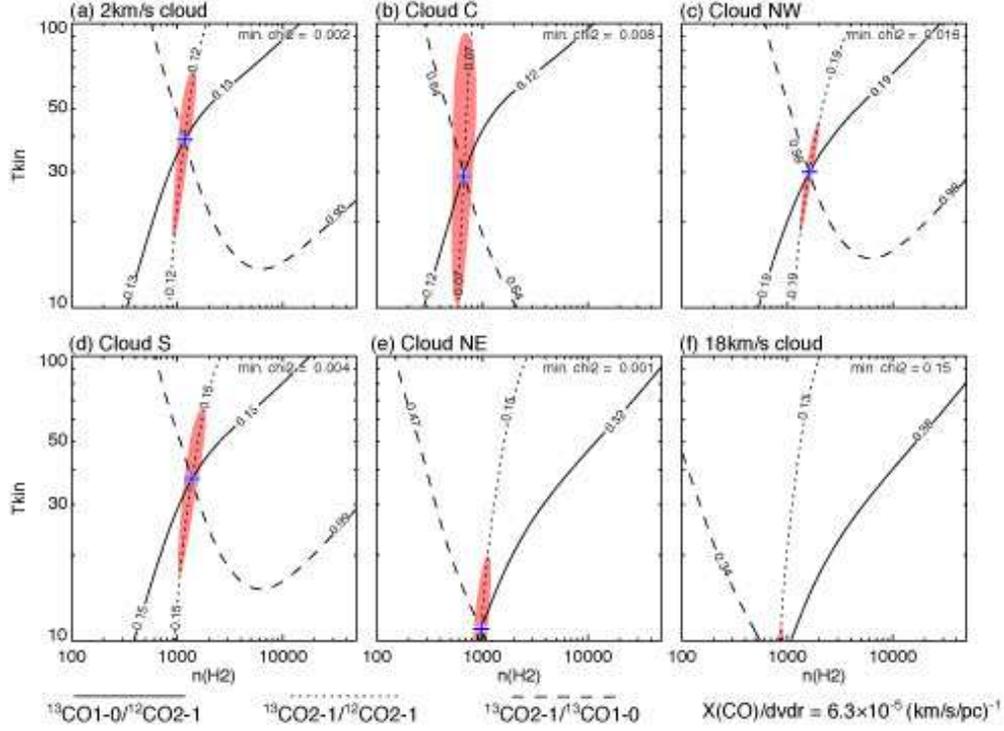}
\caption{LVG results for X(CO)/dvdr = 6.3 $\times$ 10$^{-5}$ (km s$^{-1}$ pc$^{-1}$)$^{-1}$, assuming a distance of 1.7 kpc, are shown in the density-temperature plane. Crosses denote the points of minimum $\chi^2$, and filled areas surrounding the crossses indicate $\chi^2$ of 3.84, which corresponds to the 95 \% confidence level of the $\chi^2$ distributions with 1 degree of freedom. Solid, dotted and dashed lines show $^{13}$CO($J$=1--0)/$^{12}$CO($J$=2--1), $^{13}$CO($J$=2--1)/$^{12}$CO($J$=2--1) and $^{13}$CO($J$=2--1)/$^{13}$CO($J$=1--0) intensity ratios.\label{lvg1}}
\end{figure}

\begin{figure}
\epsscale{.80}
\plotone{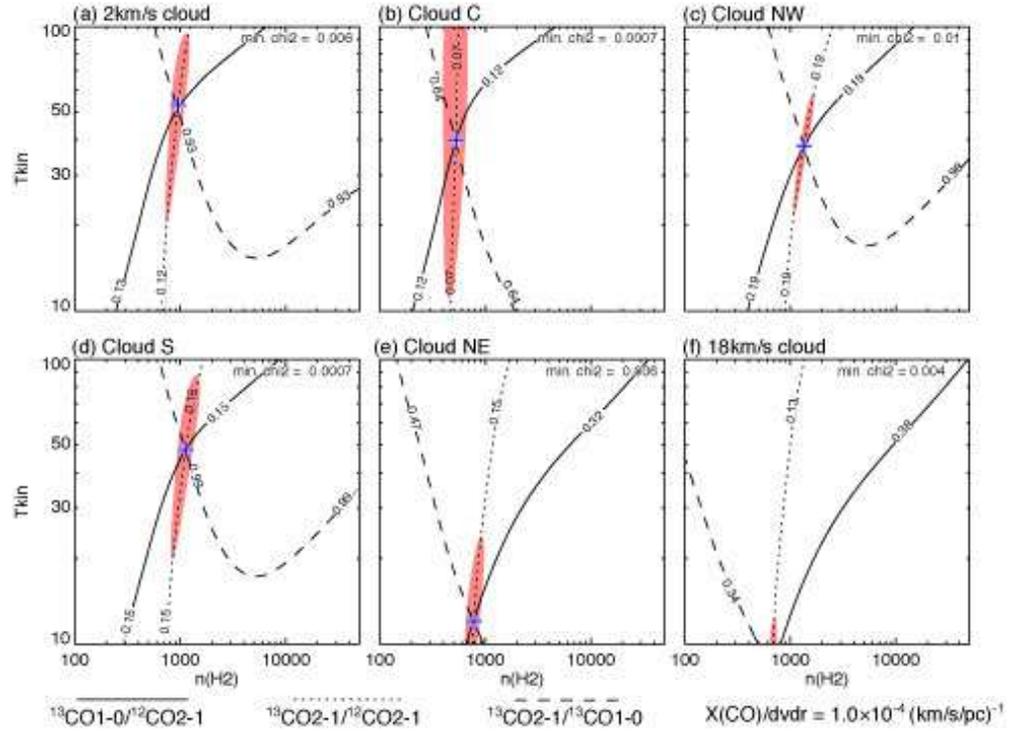}
\caption{Same as figure \ref{lvg1} for X(CO)/dvdr = 1.0 $\times$ 10$^{-4}$ (km s$^{-1}$ pc$^{-1}$)$^{-1}$, assuming a distance of 2.7 kpc. \label{lvg2}}
\end{figure}

We here assume abundance ratios of [$^{12}$CO]/[$^{13}$CO]=77 \citep{wil1994} and $X$(CO)=[$^{12}$CO]/[H$_2$]$=10^{-4}$ \citep[e.g.,][]{fre1982,leu1984}, respectively. We estimate velocity gradients of 1.6 km s$^{-1}$ pc$^{-1}$ and 1.0 km s$^{-1}$ pc$^{-1}$, and therefore $X$(CO)/($dv/dr$) of $6.3\times10^{-5}$ and $1.0\times10^{-4}$ (km s$^{-1}$ pc$^{-1}$)$^{-1}$, for distances of 1.7 kpc and 2.7 kpc, respectively. We derived $dv/dr$ by taking the average ratio between the cloud size and velocity width for the six clouds (Table \ref{cloudlist}), and then took an average of them. Figures \ref{lvg1} and \ref{lvg2} show the results of the calculations, and details of the results are summarized in Table \ref{lvglist}. The density and temperature ranges covered by the present analysis are $10^2$--$5\times10^4$ cm$^{-3}$ and 10--100 K, respectively.  In both figures, the 2 km s$^{-1}$ cloud and clouds with higher line ratios of $>$ 0.8, show kinetic temperatures of $\sim$30--50 K and densities of $\sim10^3$ cm$^3$, where slightly higher temperatures are found in the $D$=2.7 kpc case. Temperatures of $\sim$30--50 K are higher than the typical temperatures of dark clouds without a heat source, which are typically $\sim$10 K. The cloud NE shows lower temperature of $\sim$10 K. 

\section{Discussion}
\subsection{The parent cloud(s) of the cluster}

Second generation star formation is currently ongoing in M20, as evidenced by the many young stellar objects (YSOs) in the clouds \citep[see review by][]{rho2008}. On the other hand, the stars of the first generation were likely formed together with the central O star. An age of 0.3 Myrs is estimated for the HII region by considering its spatial extent \citep{cer1998} and the age of the first generation stars is thought to be less than 1 Myrs \citep[see review by][]{rho2008}. 

The distribution of the IRAS emission shown in Figures \ref{iimap}g--\ref{iimap}i is centered on the O star, and the boundary at the half peak intensity level shows an extension similar to that of the HII region \citep{cer1998}. The total infrared luminosity $L_{\rm ir}$ is given by the following equations \citep{dal2002};
\begin{eqnarray}
L_{\rm ir}  & = & 2.403\nu_{25} L_{25}-0.2454 \nu_{60}L_{60}+1.6381\nu_{100} L_{100}  \  [ L_\odot ]  \\
L_{\lambda}  & = &  4 \pi D^2 f_{\lambda} \ [ L_\odot ]
\end{eqnarray}
Here $f_\lambda$ is the flux in each of the 3 IRAS bands and $D$ is the distance to the object. $L_{\rm ir}$ for M20 is hence estimated to be $\sim2.4\times10^5$ $L_\odot$ and $6.1\times10^5$  $L_\odot$ for distances of 1.7 kpc and 2.7 kpc, respectively. \cite{dra2007} pointed out that the above equations given by \cite{dal2002} underestimate L$_{\rm ir}$ by up to 30\% within a radiation intensity $U$ range of 10--100, and applying this correction increases our estimates of $L_{\rm ir}$ to $3.4\times10^5$ $L_\odot$ and $8.6\times10^5$ $L_\odot$ for the two distances. The bolometric luminosity of the O star is $1.3$--$4.2\times10^5$ $L_\odot$ \citep{con1971,wal1973}, corresponding to 40--120 \% and 20--50 \% of the above estimates for two distances, respectively. These figures indicate that the dust luminosity is dominated by the energy of the O star, which accounts for most of the radiative energy in M20. We therefore argue that the first generation stars, including the O star, dominate the total energy release in the region and probably also make up the majority of the stellar mass contained within it. On the other hand, the stars formed in the other outer clouds NW, S and NE are much less dominant in luminosity and perhaps also in terms of the masses of individual stars.

We here discuss the heating mechanism of the warm gas in the 2 km s$^{-1}$ and clouds C, S and NW. The warm gas is distributed only within the extent of the strong infrared emission, except in the direction of the SNR (Figures \ref{iimap}g--\ref{iimap}i). It therefore seems reasonable to consider that the warm gas is heated predominantly by the O star, since the warmest regions of the 2 km s$^{-1}$ and 8 km s$^{-1}$ clouds are located within a few pc of this central source. The energy of the stellar wind from the central O star is estimated as $1.6\times10^{36}$ erg s$^{-1}$ (see section 1). The cooling rate of a cloud with a diameter of 3 pc, density of $10^3$ cm$^{-3}$ and temperature of 40 K is estimated as $\sim10^{34}$ erg s$^{-1}$ \citep{gol1978}, much lower than the energy of the stellar wind. The high temperatures of the 2 km s$^{-1}$ and 8 km s$^{-1}$ clouds can therefore be explained energetically in terms of the shock heating by the stellar winds of the central star, although the present low angular resolution is not high enough to test this possibility further by deriving a detailed temperature distribution. In either case, the physical association of the four clouds with the O star is virtually certain. 

It is therefore reasonable that both the 2 km s$^{-1}$ cloud and cloud C are candidates for the parental cloud of the O star. The two clouds share peak positions, suggesting that they are located nearly at the same position, and the velocity difference of these two clouds is $\sim$7.5 km s$^{-1}$ (Table \ref{cloudlist}), meaning that they can only move by $\sim$2 pc within the 0.3 Myr age of M20, which is consistent with the interpretation that these two clouds is located at the same position.

On the other hand, the molecular clouds at $V_{\rm lsr} \sim 18$ km s$^{-1}$ do not seem to be directly associated with the O star, because these clouds have temperatures of 10 K, lower than those of the 2 km s$^{-1}$ and 8 km s$^{-1}$ clouds. The clouds around $V_{\rm lsr} \sim$ 18 km s$^{-1}$ therefore appear to be located near M20 but are neither as close or as heated by M20 as the  2 km s$^{-1}$ and 8 km s$^{-1}$ clouds. 
Only TC3 \& TC4 cloud has a possibility to be located around M20 as close as cloud NE, NW and S, because studies by \citet{lef2000} and \citet{rho2008} indicate that it is located at the boundary of the HII region and shocks may occur there. Self absorption in the $^{12}$CO spectra possibly make it difficult to know an actual excitation condition of the cloud, and  further study in optically-thin lines is needed for better understanding.


\subsection{Possible triggering of the cluster formation}

The masses of the 2 km s$^{-1}$ cloud and cloud C, assuming distance of between 1.7 and 2.7 kpc, are estimated to $\sim0.8$--$2.0\times10^3$ $M_\odot$ and $\sim0.5$--$1.2\times10^3$ $M_\odot$, respectively, giving a total mass of about $\sim1.3$--$3.2\times10^3$ $M_\odot$. We shall first discuss the gravitational condition of these two clouds. These clouds, both peaking toward the O star, have radii of $\sim$1--2 pc and a velocity difference of $\sim$7.5 km s$^{-1}$ (Table \ref{cloudlist}). Based on these figures we estimate that a mass of $\sim$1.3--$2.6\times10^4$ $M_\odot$ is required to gravitationally bind the two clouds for an assumed separation 1--2 pc. This mass is apparently larger than the observed cloud and stellar masses, suggesting that these two clouds are not gravitationally bound to M20. 
Mechanical expansion therefore may be reasonable explanation for the velocity difference. Since M20 is a very young object with an age of 0.3 Myr, it does not yet appear to have hosted any supernova explosions, and stellar winds are the only available mechanism of driving any mechanical expansion that may be occurring around the central cluster. The total mechanical luminosity available from stellar winds of the O star over the age of the system is estimated to be $\sim1.5\times10^{49}$ erg. Assuming the velocity separation of $\sim$7.5 km s$^{-1}$ as the expanding velocity of the clouds, we obtain the total kinetic energy of the five clouds, 2 km s$^{-1}$ cloud, Cloud C, NW, S and NE, of $\sim$2.0--4.7$\times10^{48}$ erg, which corresponds to almost 15--30 \% of the energy given from the stellar winds. Numerical calculations suggest that less than a few percent of the initial wind energy is converted to neutral gas kinetic energy\citep{art2007}, and the small solid angle less than 4 $\pi$ covered by the clouds reduces the energy available to them. We therefore conclude that while mechanical expansion cannot be ruled out as the cause of the velocity separation, the energetics are uncomfortably tight. Moreover, the spatio-velocity structure of the two components, as shown in figure \ref{lv}, shows the greatest departures from the centroid velocity at around 18$^h$2$^m$00$^s$ -- the furthest projected distance from the central star. This does not fit the classical pattern of a shell-like expansion around a central source, and suggests that at least some component of the cloud velocity separation  is unrelated to their interaction with the central star

We here suggest that much of the observed motion of the clouds does not arise as a direct result of the stellar cluster but is instead systemic --- i.e. present from before the cluster's birth. Yet the 2 km s$^{-1}$ cloud and cloud C are clearly associated with M20 and in close spatial proximity to one another. We suggest that a scenario in which a cloud-cloud collision triggered the formation of the central stars is highly consistent with these observational characteristics. In this scenario the 2 km s$^{-1}$ cloud and cloud C collided each other $\sim$1 Myr ago at a relative velocity of $\sim$7.5 km s$^{-1}$ or more. The rapid collision between the two clouds strongly compressed the molecular gas and triggered the formation of the central star and surrounding first generation stars. 
In this model, the 2 km s$^{-1}$ cloud and cloud C must be moving in the opposite directions; the former is moving toward us and the latter is moving away from us. As shown in Section 3.1, the 2 km s$^{-1}$ cloud is apparently located at the front side of M20, whereas cloud C is not in the near side, suggesting that it is located either on the far side or within M20. This relative configuration is in fact consistent with the cloud-cloud collision scenario, because we must be witnessing a moment after the collision occurred Myr ago.  We would expect a reversed cloud location prior to the collision.

This scenario is similar to that discussed by \citet{fur2009} and later expanded on by \citet{oha2010}. These authors found two GMCs closely associated with the RCW 49 HII region and its exciting cluster, the super star cluster Westerlund 2. The two clouds have a velocity difference of 15 km s$^{-1}$, are not gravitationally bound to one another, and have associated kinetic energies that are too large to be explained by expansion driven by the central cluster. These characteristics are quite similar to the 2 km s$^{-1}$ and 8 km s$^{-1}$ clouds of M20, and we therefore suggest that M20 may be interpreted as a miniature of Westerlund 2.

The velocity of the M20 region as a whole (0--30 km s$^{-1}$) corresponds to the velocities of the Perseus and Scutum Arms \citep{val2008}, and there are actually many molecular clouds in the region \citep[see Figure \ref{largemap} and][]{tak2010}. Such a high density of molecular clouds implies a high cloud-cloud collision probability, providing a similar environment to that in the region of Westerlund 2. We also note that \citet{dob2001} investigated the relationship between masses of star forming molecular clouds and the infrared luminosity from these clouds, finding that molecular clouds with masses of $\sim10^3$--10$^4$ $M_\odot$ typically have infrared luminosity of up to $\sim10^4$--10$^5$ $L_\odot$ --- almost comparative to or much lower than the infrared luminosity of M20 ($\sim2.4$--$6.1\times10^5$ $L_\odot$). This fact suggests that the first generation star formation in M20 is remarkably efficient, providing further tentative support for the triggered star formati
 on mechanism suggested here.

\section{Summary}

We summarize the present works as follows;

1) A large-scale study of the molecular clouds toward the Trifid nebula, M20, has been made in the $J$=2--1 and $J$=1--0 transitions of $^{12}$CO and $^{13}$CO. These observations reveal two molecular components both peaked toward the center of M20 with LVG-derived temperatures of 30--50 K --- significantly higher than the 10 K of their surroundings.

2) The close association of the clouds with the central cluster and surrounding HII region strongly suggests that they are the parent clouds of the first generation stars in M20. The mass of each cloud is estimated to be $\sim10^3$ $M_\odot$ and their velocity separation is $\sim7.5$ km s$^{-1}$ over a length of $\sim$1--2 pc. The total stellar and molecular mass is too small by an order of magnitude to gravitationally bind the system.

3) Based on the dynamics and energetics of the system, we suggest that the formation of the first generation stars, including the central ionizing O7.5 star, was triggered by the collision between the two clouds on a short time scale of $\sim$1 Myr. This is a second case alongside that of the super star cluster Westerlund 2, for which a similar collision-triggered formation mechanism has been suggested. 

\acknowledgments
NANTEN2 is an international collaboration of ten universities, Nagoya University, Osaka Prefecture University, University of Cologne, University of Bonn, Seoul National University, University of Chile, University of New South Wales, Macquarie University, University of Sydney and Zurich Technical University. NANTEN was operated based on a mutual agreement between Nagoya University and the Carnegie Institution of Washington. We also acknowledge that the operation of NANTEN can be realized by contributions from many Japanese public donators and companies. The work is financially supported by a grant-in-aid for Scientific Research (KAKENHI, no. 15071203, no. 21253003, and no. 20244014) from MEXT (the Ministry of Education, Culture, Sports, Science and Technology of Japan) and JSPS (Japan Society for the Promotion of Science) as well as JSPS core-to- core program (no. 17004).  This research was supported by the grant-in-aid for Nagoya University Global COE Program, ¡ÈQuest for Fun
 damental Principles in the Universe: from Particles to the Solar System and the Cosmos,¡É from MEXT. Also, the work makes use of archive data acquired with IRAS and Spitzer data sets gained with Infrared Processing and Analysis Center (IPAC).

\clearpage

\clearpage

\clearpage

\begin{table}
\begin{center}
\tabletypesize{\scriptsize}
\caption{Physical properties of the clouds.\label{cloudlist}}
\begin{tabular}{ccccccc}
\tableline\tableline
 \multirow{2}{*}{Name} & Ra(J2000) & Dec(J2000) & $V_{\rm lsr}$& $\Delta V$ & $r$\tablenotemark{\dagger} & $M$(H$_2$)\tablenotemark{\dagger} \\
	  & ($^h \ ^s \ ^m$) & ($^\circ \ \arcmin \ \arcsec$) & (km s$^{-1}$) &(km s$^{-1}$)  & (pc) & ($\times$10$^3$ $M_\odot$) \\
\tableline
2 km s$^{-1}$ cloud 	& 18:02:19.50  	& $-$23:01:40.4 	& 1.4		& 4.0 	& 1.2/2.0 	& 0.9/2.0\\
cloud C 			& 18:02:18.36 	& $-$23:01:45.3 	& 8.9		& 3.6 	& 0.9/1.5 	& 0.6/1.2\\
cloud NW 			& 18:02:04.31 	& $-$22:56:48.8 	& 10.1	& 3.3 	& 1.2/1.9 	& 0.9/2.3\\
cloud S 			& 18:02:25.22  	& $-$23:05:42.9 	& 7.5		& 3.7 	& 0.8/1.2 	& 0.6/1.0\\
cloud NE 			& 18:02:51.35  	& $-$22:58.31.7		& 9.9		& 2.6 	& 1.2/2.0 	& 0.7/1.8\\
18 km s$^{-1}$ cloud	& 18:02:32.31  	& $-$23:03:25.1		& 18.2	& 6.4 	& 2.3/3.7 	& 5.6/14.1\\
TC3 \& TC4 cloud  	& 18:02:16.30	& $-$23:05:25.1		& ---\tablenotemark{\ddagger} 	& ---\tablenotemark{\ddagger} 		& 1.0/1.6	& 0.8/2.2	\\

\tableline
\end{tabular}
\tablenotetext{\dagger}{Sizes and masses of the clouds were calculated with the both distances of 1.7kpc/2.7kpc. }
\tablenotetext{\ddagger}{$V_{\rm lsr}$ and $\Delta V$ of TC3 \& TC4 cloud can not be measured due to self absorption in the spectrum.}
\tablecomments{Column (1): Name of cloud.  (2--3): Peak position of cloud. (4): Velocity of cloud. (5): Velocity dispersion (FWHM). (6): Radius of cloud. (7): Cloud mass derived using an X-factor of 2.0 $\times$ 10$^{20}$cm$^{-2}$ (K km s$^{-1}$)$^{-1}$.}
\end{center}
\end{table}

\clearpage

\begin{table}
\begin{center}
\caption{LVG results.\label{lvglist}}
\begin{tabular}{cccccccc}
\tableline\tableline
 \multirow{2}{*}{Name} & \multicolumn{3}{c}{$D$ = 1.7 kpc} && \multicolumn{3}{c}{$D$ = 2.7 kpc} \\
\cline{2-4} \cline{6-8}
	&  $n$(H$_2$) (cm$^{-3}$) & $T_{\rm{k}}$ (K) & min. $\chi^2$ && $n$(H$_2$) (cm$^{-3}$) & $T_{\rm{k}}$ (K) & min. $\chi^2$ \\
\tableline

2 km s$^{-1}$ cloud 	& $1.2^{+0.3}_{-0.3}\times10^3$ 	& 39$^{+28}_{-22}$	& 0.002 &
				& $1.0^{+0.2}_{-0.3}\times10^3$ 	& 53$^{+40}_{-32}$	& 0.006 \\
				
cloud C 			& $0.7^{+0.2}_{-0.2}\times10^3$ 	& 29$^{+62}_{-19}$ 	& 0.008 &
				& $0.5^{+0.2}_{-0.1}\times10^3$ 	& $> 40_{-29}$ 	& 0.0007 \\
				
cloud NW			& $1.6^{+0.4}_{-0.5}\times10^3$ 	& 30$^{+14}_{-13}$	& 0.02 &
				& $1.4^{+0.3}_{-0.4}\times10^3$ 	& 38$^{+19}_{-17}$	& 0.01 \\
				
cloud S			& $1.4^{+0.5}_{-0.4}\times10^3$ 	& 37$^{+27}_{-20}$	& 0.004 &
				& $1.1^{+0.5}_{-0.3}\times10^3$ 	& 48$^{+40}_{-28}$	& 0.004 \\
				
cloud NE 			& $1.0^{+0.1}_{-0.2}\times10^3$ 	& $< 11^{+9}$		& 0.001 &
				& $0.8^{+0.1}_{-0.2}\times10^3$ 	& $< 12^{+11}$		& 0.006 \\
				
18 km s$^{-1}$ 		& $0.9^{+0.1}_{-0.1}\times10^3$		& $< 10^{+1}$	 	& 0.15 &
				& $0.7^{+0.1}_{-0.1}\times10^3$		& $< 10^{+2}$	 	& 0.08 \\

\tableline
\end{tabular}
\tablecomments{Column (1): Name of cloud.  (2--4): LVG result for $X$(CO)($dv/dr$) = $6.3\times10^{-5}$ (km s$^{-1}$ pc$^{-1}$) at a distance $D$ of 1.7 kpc. (5--7): LVG result for $X$(CO)($dv/dr$) = $1.0\times10^{-4}$ (km s$^{-1}$ pc$^{-1}$) at a distance $D$ of 2.7 kpc. (2, 5): Number density of H$_2$. (3, 6): Kinetic temperature. (4, 7) minimum $\chi^2$.}
\end{center}
\end{table}







\begin{thebibliography}{}

\bibitem[Arthur(2007)]{art2007}Arthur,~S.~J., 2007, in IAU Symp. 250, Massive Stars as Cosmic Engines, ed. F. Bresolin, P. A. Crowther, \& J. Puls (Cambridge: Cambridge Univ. Press), 355
\bibitem[Arikawa et al.(1999)]{ari1999}Arikawa,~Y., Tatematsu,~K., Sekimoto,~Y., \& Takahashi,~T., 1999, \pasj, 51, 7
\bibitem[Ascenso et al.(2007)]{asc2007} Ascenso,~J., Alves,~J., Beletsky,~Y., \& Lago,~M.~T.~V.~T., 2007, \aap, 466, 137
\bibitem[Cambr\'esy et al.(2011)]{cam2011}Cambr\'esy,~L., Rho,~J., Marshall,~D.~J., \& Reach,~W.~T., \aap, 527, 141
\bibitem[Cernicharo et al.(1998)]{cer1998}Cernicharo,~J. et al., 1998, Science, 282, 462
\bibitem[Chaisson \& Willson(1975)]{cha1975}Chaisson,~E.~J. \& Willson,~R.~F., 1975, \apj, 199, 647
\bibitem[Conti \& Alschuler(1971)]{con1971}Conti,~P.~S., \& Alschuler,~W.~R., 1971, \apj, 170, 325
\bibitem[Dawson et al.(2011)]{daw2011} Dawson,~J.~R., McClure-Griffiths,~N.~M., Kawamura,~A., Mizuno,~N., Onishi,~T., Mizuno,~A., \&  Fukui,~Y.,  2011, \apj, 728, 127
\bibitem[Dale \& Helou(2002)]{dal2002}Dale,~A.~D., \& Helou,~G., 2002, \apj, 576, 159
\bibitem[Dobashi et al.(2001)]{dob2001}Dobashi,~K., Yonekura,~Y., Matsumoto,~T., Momose,~M., Sato,~F., Bernard,~J., \& Ogawa,~H., 2001, \pasj, 53, 85
\bibitem[Draine \& Li(2007)]{dra2007}Draine,~B.~T., \& Li,~A., 2007, \apj, 657, 810
\bibitem[Frerking, Langer \& Wilson(1982)]{fre1982}Frerking,~M.~A., Langer,~W.~D., \& Wilson,~R.~W., 1982, \apj, 262, 590
\bibitem[Furukawa et al.(2009)]{fur2009} Furukawa,~N., Dawson,~J.~R., Ohama,~A., Kawamura,~A., Mizuno,~N., Onishi,~T., \& Fukui,~Y., 2009,  \apj, 696L, 115 
\bibitem[Goldreich \& Kwan(1974)]{gol1974} Goldreich,~P., \& Kwan,~J., 1974, \apj, 189, 441
\bibitem[Goldsmith \& Langer(1978)]{gol1978} Glodsmith,~P.~F., \& Langer,~W.~D., 1978, \apj, 222, 881
\bibitem[Herbig(1957)]{her1957} Herbig,~G.~H., 1957, \apj, 125, 654
\bibitem[Hester et al.(2004)]{hes2004} Hester,~J.~J., Desch,~S.~J., Healy,~K.~R., \& Leshin,~L.~A., 1990, Science, 304, 1116
\bibitem[Higuchi et al.(2010)]{hig2010}Higuchi,~A.~E., Kurono,~Y., Saito,~M., \& Kawabe, R., 2010, \apj, 719, 1813
\bibitem[Howarth \& Prinja(1989)]{how1989} Howarth,~I.~D. \& Prinja,~R.~K., 1989, 69, 527
\bibitem[Lefloch \& Cernicharo(2000)]{lef2000} Lefloch,~B. \& Cernicharo,~J., 2000, \apj, 545, 340
\bibitem[Lefloch et al.(2001)]{lef2001} Lefloch,~B., Cernicharo,~J., Cesarsky,~D., Demyk, K., \& Rodr\'iguez, L. F., 2001, \aap, 368, 13
\bibitem[Lefloch et al.(2002)]{lef2002} Lefloch,~B., Cernicharo,~J., Rodr\'iguez, L. F., Miville-Desch$\hat{\rm e}$nes,~M.~A., Cesarsky,~D., \& Heras,~A., 2002, \apj, 581, 335
\bibitem[Lefloch et al.(2008)]{lef2008} Lefloch,~B., Cernicharo,~J., \& Pardo,~J.~R., 2008, \aap, 489, 157
\bibitem[Leung, Herbst \& Huebner(1984)]{leu1984}Leung,~C.~M., Herbst,~E., \& Huebner,~W.~F., 1984, \apjs, 56, 231
\bibitem[Lefung \& Liszt(1976)]{leu1976}Leung,~C.~-M., \& Liszt,~H.~S., 1976, \apj, 208, 732
\bibitem[Lynds, Canzian \& O'neil(1985)]{lyn1985}Lynds,~B.~T., Canzian,~B.~J. \& O'neil,~E.~J.,~Jr., 1985, \apj, 288, 164
\bibitem[Ohama et al.(2010)]{oha2010} Ohama,~A. et al., 2010, \apj, 709, 975
\bibitem[Ogawa et al.(1990)]{oga1990}Ogawa,~H., Mizuno,~A., Hoko,~H., Ishikawa,~H., \& Fukui, Y., 1990, Int. J. Infrared and Millimeter Waves, 11, 717
\bibitem[Ogura \& Ishida(1975)]{ogu1975}Ogura,~K., \& Ishida,~K., 1975, \pasj, 27, 119
\bibitem[Prinja, Barlow \& Howarth(1990)]{pri1990}Prinja,~R.~K., Barlow,~M.~J., Howarth,~I.~D., 1990, \apj, 361, 607
\bibitem[Rauw et al.(2007)]{rau2007}Rauw,~G. et al., 2007, \araa, 463, 981
\bibitem[Rho et al.(2001)]{rho2001} Rho,~J., Corcoran,~M.~F., Chu,~Y., \& Reach,~W.~T., 2001, \apj, 562, 446
\bibitem[Rho et al.(2004)]{rho2004} Rho,~J., Ram\'irez,~S.~V., Corcoran,~M.F., Hamaguchi,~K., \& Lefloch,~B., 2004, \apj, 607, 904
\bibitem[Rho et al.(2006)]{rho2006} Rho,~J., Reach,~W.~T., Lefloch,~B., \& Fazio,~G.~G.,  2006, \apj, 643, 965
\bibitem[Rho et al.(2008)]{rho2008} Rho,~J., Lefloch,~B., Reach,~W.~T., \& Cernicharo,~J., 2008, in Handbook of Star Forming Regions, vol II, ASP Conf. Ser., 5, 509
\bibitem[Strong et al.(1988)]{str1988}Strong,~A.~W. et al., 1988, \aap,207,1
\bibitem[Takeuchi et al.(2010)]{tak2010}Takeuchi,~T. et al., 2010, \pasj, 62, 557
\bibitem[Vall\'ee(2008)]{val2008} Vall\'ee,~J.~P., 2008, \aj, 135, 1301
\bibitem[Walborn(1973)]{wal1973} Walborn,~N.~R., 1973, \aj, 78, 1067
\bibitem[Wilson \& Rood(1994)]{wil1994}WIlson,~T.~L., \& Rood,~R., 1994, \araa, 32, 191
\bibitem[Wilson et al.(2000)]{wil2000}Wilson,~C.~ D., Scoville,~N., Madden,~S.~C., Charmandaris,~V., 2000, \apj, 542, 120
\bibitem[White(1977)]{whi1977}White,~R.~E., 1977, \apj,211,744
\bibitem[Yamaguchi et al.(2001)]{yam2001}Yamaguchi,~R., Mizuno,~N., Onishi,~T., Mizuno,~A., \& Fukui,~Y., 2001, \pasj, 53, 959
\bibitem[Yusef-Zadeh, Biretta \& Geballe(2005)]{yus2005} Yusef-Zadeh,~F., Biretta,~J., \& Geballe,~T.~R. 2005, \aj, 130, 1171
\bibitem[Zinnecker \& Yorke(2007)]{zin2007}Zinnecker,~H., \&  Yorke,~H.~W., 2007, \araa, 45, 481

\end{thebibliography}
\end{document}